\documentclass[twocolumn,english,aps,prl,floatfix,twocolumn,reprint,amsmath,amssymb,superscriptaddress]{revtex4-2}
\usepackage[T1]{fontenc}
\usepackage[utf8]{inputenc}
\usepackage{color}
\usepackage{babel}
\usepackage{amsmath}
\usepackage{amssymb}
\usepackage{graphicx}
\usepackage{esint}
\usepackage[unicode=true,
 bookmarks=false,
 breaklinks=false,pdfborder={0 0 1},backref=section,colorlinks=true]
 {hyperref}
\hypersetup{
 letterpaper=true,pdfstartview=FitV,linkcolor=blue,citecolor=blue,urlcolor=blue}

\makeatletter
\usepackage{babel}
\usepackage{mathrsfs}
\usepackage{color}
\usepackage{bm}
\usepackage{xspace}
\usepackage{epstopdf}
\usepackage{dcolumn}
\usepackage{longtable}
\usepackage{multirow}

\makeatother

\begin{document}
\title{3D Topological Plasmons in Weyl Semimetals}
\author{Furu Zhang}
\author{Yang Gao}
\affiliation{Institute of Applied Physics and Computational Mathematics, Beijing 100088, China}
\author{Wei Zhang}
\email{zhang_wei@iapcm.ac.cn}

\affiliation{Institute of Applied Physics and Computational Mathematics, Beijing 100088, China}
\begin{abstract}
We systematically investigate the properties of bulk, surface and edge plasmons in Weyl semimetals in presence of a magnetic field. It is found that unidirectional plasmons with different properties exist on different surfaces, which is in consistent with the nontrivial topology of the three-dimensional (3D) bulk plasmons. These novel plasmons possess momentum-location lock and may travel between surfaces. The anomalous Hall conductivity brings about abundant anisotropic plasmon dispersions, from linear to parabolic or even hyperbolic bands. With the help of a semiclassical picture for the formation of the Weyl orbits, we point out the Fermi-arc plasmons at opposite surfaces can make up another unique 3D topological plasmon. Furthermore, there is a gapless unidirectional edge plasmon protected by the topology whose direction and position can be controlled by external field. Our work thus uncovers topological features of Weyl plasmons, which may have important applications in photoelectric devices based on chiral/topological plasmonics.
\end{abstract}
\maketitle
$Introduction.$---Weyl semimetal is a 3D analogue of graphene with linear low energy excitations, possessing topologically protected Weyl nodes in the bulk and Fermi arcs on surfaces \cite{Qi 2013,Burkov 2016,Felser 2017}. Several theoretical predictions and experimental observations have been proposed for realizing Weyl semimetals in real materials \cite{Wan 2011,Xu 2011,Weng 2015,Hasan 2015,Chen 2015,Ding 2015,Vishwanath 2018}. Because of the Berry flux between Weyl nodes, Weyl semimetal owns a 3D anomalous Hall effect and the intrinsic Hall conductivity is proportional to the distance between Weyl nodes \cite{Y. Ran 2011,Burkov 2014}. In the presence of a perpendicular magnetic field, the Fermi arcs and chiral zeroth Landau levels will form closed Weyl orbits \cite{Vishwanath 2014,Analytis 2016,C. Zhang 2017}, giving rise to an unique 3D quantum Hall effect \cite{X. C. Xie 2017,C. Zhang 2019}. Whereas, the relationship between Weyl orbits and the Berry curvature of Fermi arcs is not known, nor the velocity of the electron tunneling between surfaces. Further more, what's the impact of Weyl orbits on the collective excitations? Our studies in this letter make a step in the research on these interesting topics.

Weyl semimetal can serve as a potential optical material with electromagnetic response described by a topological Chern-Simon term. As a fundamental optical property, the collective excitation of Weyl electrons has been studied extensively. The frequency of the bulk Weyl plasmons depends on the position of Fermi surface and is related to the chiral anomaly \cite{J. Zhou 2015,Das Sarma 2015}. Due to the effective magnetic field in momentum space, the surface Weyl plasmons are similar to magnetoplasmons in ordinary metals \cite{Das Sarma 2016,Lo 2018}. The Fermi-arc plasmons are chiral and have unusual band structures \cite{J. C. W. Song 2017,G. M. Andolina 2018,F. Adinehvand 2019}. In the presence of a magnetic field, plasmons in Weyl semimetals earn more interesting optical properties, such as chiral electric separation, coupling-induced transparency and nonclassical density response \cite{Panfilov 2014,Sukhachov 2017,Belyanin 2018}. However, there is no discussion about nontrivial topology of Weyl plasmons, nor influence of the ``wormhole'' tunneling on the plasmon transport property.

In this letter, we point out that the bulk Weyl magnetoplamons and the Fermi-arc magnetoplasmons over opposite surfaces are both 3D topological plasmons. In consistent with the nontrivial topology of the bulk (surface) states, there are novel unidirectional surface (edge) plasmons whose direction and dispersion can be controlled by the external field. These chiral surface plasmons own momentum-location lock and may travel between surfaces when reaching boundaries. The anomalous Hall conductivity could influence the magnetoplasmon dispersion greatly, resulting in linear (gapless), parabolic or even hyperbolic (gapped) bands. Strong confinement of electromagnetic (EM) field associated with the Fermi-arc plasmons can be achieved. We also propose a semiclassical picture of the formation of Weyl orbits, connecting the opposite Fermi-arc plasmons into a whole one. Our work thus sheds some new light on the electron dynamics and collective excitations of Weyl fermions and proposes potential applications of Weyl semimetals in 3D chiral and topological plasmonics.

$Model\ of\ Weyl\ semimetals.$---To illustrate the transport properties of plasmons in presence of a uniform magnetic field $\boldsymbol{B}=(0,B,0)$, we start from a minimal model of Weyl semimetals \cite{Okugawa 2014}:

\begin{equation}
H=A(k_{x}\sigma_{x}+k_{y}\sigma_{y})+M(b^{2}-k^{2})\sigma_{z}+D_{1}k_{y}^{2}+D_{2}k_{\parallel}^{2},
\end{equation}
which breaks the time-reversal symmetry and hosts a pair of Weyl nodes at $(0,0,\pm b).$ $\boldsymbol{k}=(k_{x},k_{y},k_{z})$ and $k_{\parallel}^{2}=k_{x}^{2}+k_{z}^{2}$. The effective model for the Fermi arc at the top ($\tau=1$) or bottom ($\tau=-1$) surface reads \cite{X. C. Xie 2017}:

\begin{equation}
h_{\tau}=\tau vk_{x}+(D_{2}-D_{1})k_{\parallel}^{2}+D_{1}b^{2}\text{,}
\end{equation}
where $v=A\sqrt{M^{2}-D_{1}^{2}}/M$. The anisotropic parameters $D_{1}$ and $D_{2}$ make the Fermi arc a two-dimension (2D) dispersion. The Fermi level lying at the Weyl nodes is $E_{F}=D_{2}b^{2}$. At the top surface, the Fermi-arc electrons ($k_{x}>0$) have a positive velocity ($\partial h_{\tau}/\partial k_{x}$) propagating along the $x$ direction; oppositely, the electron velocity is negative at the bottom surface ($k_{x}<0$).

The Weyl semimetal may have a large anomalous Hall conductivity \cite{Z. Fang 2003,Large 2018}. For any $k_{z}\in(-b,b)$, there is a well defined Chern number $C_{k_{z}}=sgn(M)$ \cite{Y. Ran 2011} in the conduction band which is related to the existence of the Fermi arcs. As a result, the anomalous Hall conductivity is proportional to the distance between Weyl nodes: $\sigma_{H}=\sigma_{yx}$$=sgn(M)\frac{e^{2}}{2\pi h}2b$ \cite{Y. Ran 2011,Burkov 2014}. With an applied magnetic field, the bulk conductivity $\boldsymbol{\hat{\sigma}}$ and the surface conductivity $\boldsymbol{\sigma}^{s}$ will take changes accordingly (see Sec. S1 of \cite{Supplemental}).

$Chiral\ Surface\ Plasmons.$---Below, we will give a detailed investigation of the surface plasmons in Weyl semimetals, and we shall focus on the undoped case. We start from the electrodynamic equations of the Weyl electrons. We assume the thickness of the slab $L$ is very large and the electric potential near the top surface ($y=0$) takes the form as $\phi(x,y<0,z)=\phi_{0}e^{ky}e^{i(q_{z}z+q_{x}x-\omega t)}$ (discussions about plasmons on the right and front surfaces can be seen in Sec. S3 and S4 of \cite{Supplemental}). Accordingly, the electric field $\boldsymbol{E}=-\nabla\phi$, the charge density $\rho$ and the current density $\boldsymbol{j}$ hold the same form. From the Poisson equation $\nabla^{2}\phi=-\rho/\varepsilon_{0}\epsilon\Theta(-y),$ the charge conservation equation $\frac{\partial\rho}{\partial t}+\nabla\cdot\boldsymbol{j}=0,$ and the microscopic Ohm's law $\boldsymbol{j}=\hat{\boldsymbol{\sigma}}\boldsymbol{E},$ we obtain the bulk relationship (see Sec. S2.1 of \cite{Supplemental}):

\begin{equation}
\varepsilon_{0}\epsilon(q^{2}-k^{2})=\frac{\eta k(q_{z}+i\eta q_{x})\sigma_{H}}{i\omega(1-\eta^{2})},
\end{equation}
where $q=\sqrt{q_{x}^{2}+q_{z}^{2}}$ and $\eta=\omega_{c}/\omega$ with a cyclotron frequency $\omega_{c}$. $\varepsilon_{0}$ is the permittivity of vacuum, $\epsilon=13$ is the static dielectric constant of the medium \Citep{Sushkov 2015}. Instead of the topological Chern-Simon term modifying the Maxwell’s equations, the information of the topology of Weyl electrons here is completely described by the anomalous Hall conductivity $\sigma_{H}$.

At the top surface, the boundary condition gives the constitutive relation which determines the Fermi-arc plasmon dispersions:

\begin{equation}
\varepsilon_{0}(q+\epsilon k)=\frac{q_{x}\sigma_{H}}{\omega}+\frac{q_{z}^{2}\sigma_{zz}^{s}+q_{x}^{2}\sigma_{xx}^{s}+iq_{z}q_{x}\eta(\sigma_{zz}^{s}-\sigma_{xx}^{s})}{i\omega(1-\eta^{2})},
\end{equation}
where $\sigma_{jj}^{s}$ is the surface conductivity which can be calculated by the Hamiltonian Eq.(2) of the surface states or by defining from the bulk conductivity as $\sigma^{s}=\sigma_{bulk}L$ \cite{X. C. Xie 2017}. Numerical calculation and theoretical derivation both suggest that $\sigma_{jj}^{s}$$=$$\alpha\varepsilon_{0}\tilde{\sigma}_{jj}^{s}$$\approx$$\alpha\varepsilon_{0}\frac{iD_{j}}{\omega-vq_{x}}$ where $\alpha=e^{2}/h\varepsilon_{0}$ and $D_{j}$ is the Drude weight in the ac conductivity (see Sec. S2.3-S2.4 of \cite{Supplemental}).

For the intrinsic case $\omega_{c}=0$, if $\sigma_{H}\rightarrow0$ and $\omega\gg vq_{x}$, one obtains the traditional 2D plasmon whose frequency is proportional to $\sqrt{q}$: $\omega$$=$$\sqrt{\alpha D_{0}q/2\epsilon_{eff}}$, where $D_{z}=$$D_{x}=$$D_{0}$, $\epsilon_{eff}$$=$$(1+\epsilon)/2$. In general, Eq.(4) indicates anisotropic Fermi-arc plasmons in accordance with the anisotropy in the Fermi-arc dispersion. For the $z$ direction, $\omega$$=$$\sqrt{\alpha D_{z}q_{z}/2\epsilon_{eff}}$; for the $x$ direction, in the low frequency range $\omega\ll\alpha\tilde{\sigma}_{H}$ where $\tilde{\sigma}_{H}=\sigma_{H}/\alpha\varepsilon_{0}$, one can get a linear plasmon $\omega$$=$$(v-D_{x}/\tilde{\sigma}_{H})q_{x}$, which is consistent with the result utilizing 3D dielectric function \cite{F. Adinehvand 2019}. The anisotropic plasmon dispersion results in different EM responses for external fields with different polarization. Unlike the case with polarization along $z$ direction, the plasmonic modes with $q_{z}=0$ can not be excited directly due to $v/\hbar$$<$$c$, just as the usual surface plasmon at the interface between metal and vacuum.

When the external magnetic field is applied, a gap will be opened in the plasmon dispersion \Citep{Fetter 1985,D. Jin 2016}. If $\sigma_{H}\rightarrow0$, $\omega\gg vq_{x}$, one can get the traditional 2D magnetoplasmon $\omega=\sqrt{\frac{\alpha D_{0}q}{2\epsilon_{eff}}+\omega_{c}^{2}}$. If $\tilde{\sigma}_{H}\gg\omega/\alpha$ and $q_{x}=0$, numerical calculation suggests that $Re[k]\ll q_{z}$. In the long wave limit $q_{z}\rightarrow0$, Eq.(4) gives a novel gapped mode $\omega=$$\sqrt{\alpha D_{z}|q_{z}|+\omega_{c}^{2}}$, which differs from the traditional magnetoplasmon.

In the $x$ direction, the magnetic field and the anomalous Hall conductivity bring about abundant plasmon dispersions. In the long wave limit $q_{x}\rightarrow0$, we have

\begin{equation}
\omega=(v-D_{x}/\beta\tilde{\sigma}_{H})q_{x},
\end{equation}
as plotted in Fig.1(a). When $\omega_{c}=0$, it turns back into the intrinsic mode with $\beta=1$. While in the limit $\omega_{c}\rightarrow+\infty$, we have $\beta\rightarrow+\infty$ and $\omega=vq_{x}$ (see Sec. S2.5 of \citep{Supplemental}). It is also a gapless linear magnetoplasmon. Because $\omega\geq0$, one obtains $q_{x}\geq0$ implying that it propagates unidirectionally only along the positive $x$ direction. These properties are significantly different from the traditional surface magnetoplasmons which are gapped. In the next section we will show that it is a surface counterpart of the topological bulk plasmon.

When $\omega\gtrsim\omega_{c}$, from Eq.(3) we have $k\approx0$. If $\tilde{\sigma}_{H}\gg\omega/\alpha$, from Eq.(4) one can get that ($q_{z}=0$):
\begin{equation}
(1-\frac{\omega_{c}^{2}}{\omega^{2}})(1-\frac{\omega}{vq_{x}})=\frac{D_{x}}{v\tilde{\sigma}_{H}},
\end{equation}
which gives an unusual plasmon with a hyperbolic band. Let the right side be zero ($D_{x}=0$), we get the equations of two asymptotes $\omega_{1}=vq_{x}$ and $\omega_{2}=\omega_{c}$. In the limit $\omega/\omega_{c}\rightarrow+\infty$, Eq.(6) gives the intrinsic dispersion that $\omega=(v-D_{x}/\tilde{\sigma}_{H})q_{x}$. The exact solution is plotted in Fig.1(a). From the dispersion relationship, one can obtain $q_{x}\geq\frac{\omega_{c}}{v}(1+\sqrt{2D_{x}/v\tilde{\sigma}_{H}})\equiv\bar{q}$ for the branch $\omega\geq\omega_{c}$. Then in the real space there is a confinement of the EM field associated with the Fermi-arc plasmon. The maximum confinement length $1/\bar{q}$ can be tuned by the external magnetic field. This is quite different from the case in intrinsic plasmons or traditional magnetoplasmons.

It is worth noting that, in Fig.1(a), the plasmon momentum $q_{x}$ is locked with its surface location. This is a result due to the particular way of electron transitions in Fermi arcs. For the top Fermi arc, the electrons with dispersion $h_{\tau=1}\approx vk_{x}$ can only absorb the photons whose momentum is positive; but for the bottom surface, the photons' momentum must be negative. In the next section we will show that it's also a result from the nontrivial bulk topology.

\begin{figure}
\noindent \begin{centering}
\includegraphics[scale=0.32]{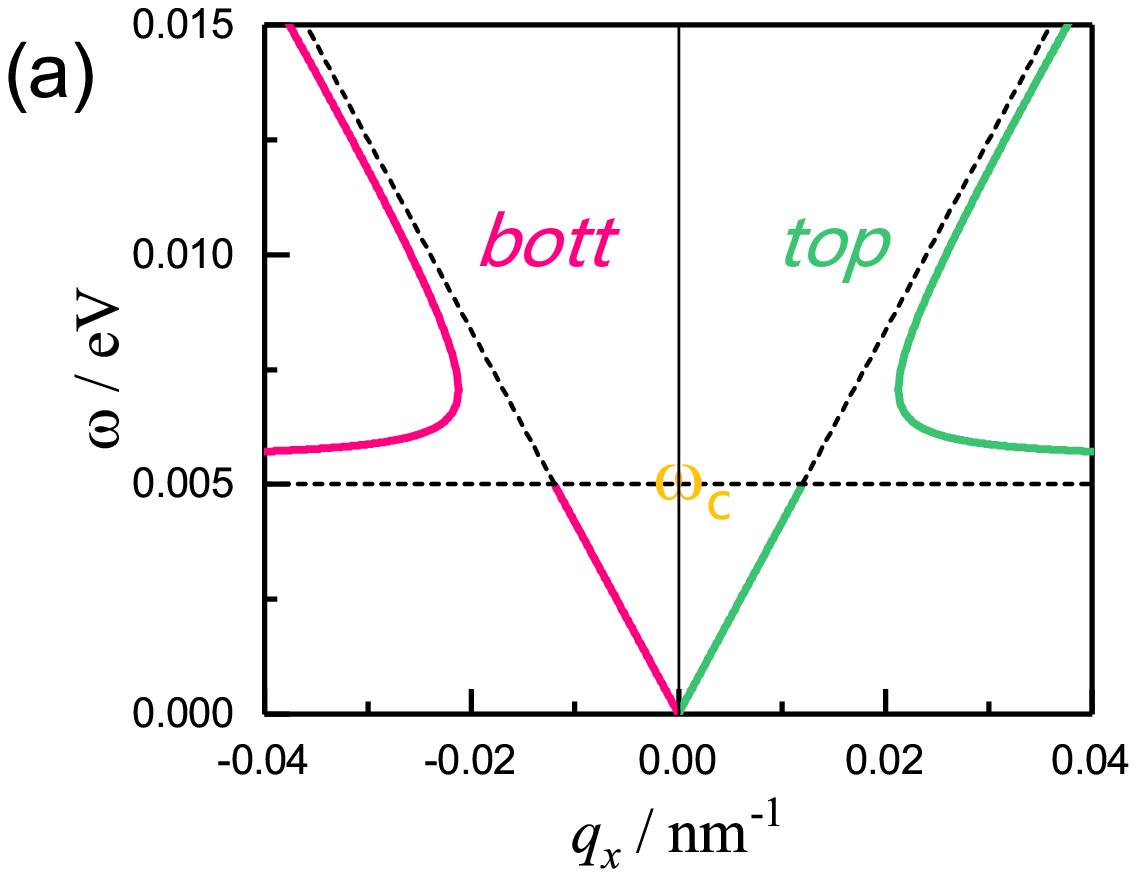}\ \ \ \ \ \includegraphics[scale=0.4]{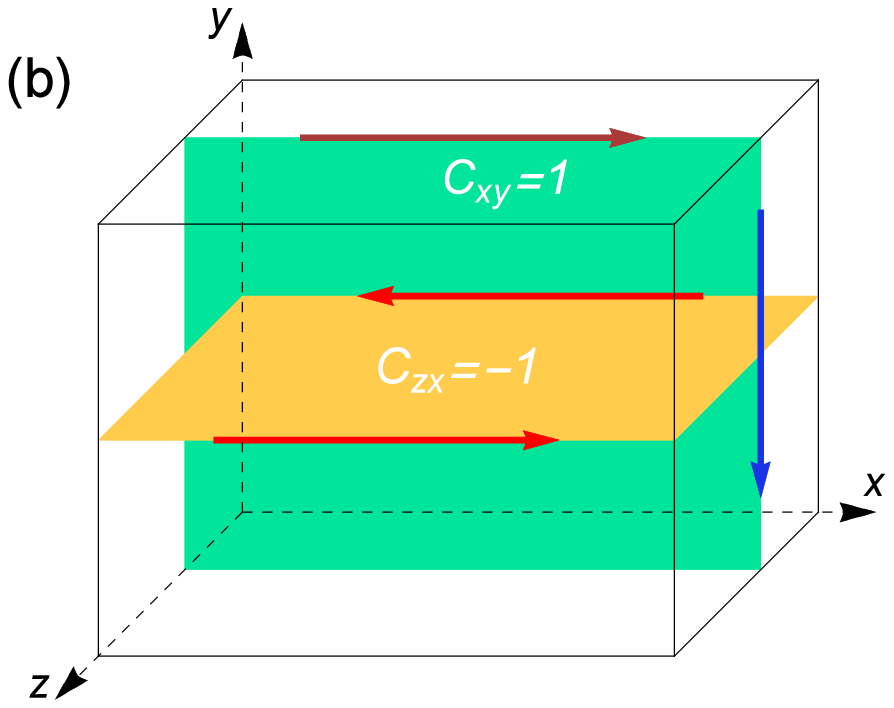}
\par\end{centering}
\caption{(a) Dispersion of Weyl surface plasmons on the top and bottom surfaces. $D_{x}=0.05$eV, $q_{z}=0$, $\hbar\omega_{c}=5$meV, $v=0.5\mathrm{eV\cdot nm}$, $\tilde{\sigma}_{H}=0.6$. The dashed lines denote the dispersions of $\omega=\pm(v-D_{x}/\tilde{\sigma}_{H})q_{x}$ and $\omega=\omega_{c}$. (b) Schematic diagrams of the propagation of the surface plasmons and the corresponding bulk topology. See Sec. S3 and S4 of \citep{Supplemental} for discussions about plasmons on the right and front surfaces.}
\end{figure}

$Topological\ bulk\ plasmons.$---Below we will point out that the Weyl semimetal can host a 3D topological plasmon with help of the magnetic field, and the unidirectional surface plasmons can be classified according to the topology of the bulk.

We come from the set that $\phi(\boldsymbol{r})$$=$$\phi_{0}$$e^{i(q_{x}x+q_{y}y+q_{z}z-\omega t)}$ and $\rho(\boldsymbol{r})$$=$$\rho_{0}$$e^{i(q_{x}x+q_{y}y+q_{z}z-\omega t)}$. The electrodynamic equations give the constitutive relation that

\begin{equation}
\varepsilon_{0}\epsilon q^{2}=\frac{\omega_{c}q_{y}\sigma_{H}(q_{z}+i\eta q_{x})}{\omega^{2}-\omega_{c}^{2}}.
\end{equation}
Here $q=\sqrt{q_{x}^{2}+q_{y}^{2}+q_{z}^{2}}$. Near the cyclotron frequency $\omega_{c}$ where $\eta\rightarrow1$ we have

\begin{equation}
\omega=\sqrt{\frac{\omega_{c}q_{y}\sigma_{H}(q_{z}+iq_{x})}{\varepsilon_{0}\epsilon q^{2}}+\omega_{c}^{2}}.
\end{equation}
Because the Fermi level is on the Weyl nodes, we have $\sigma_{ii}=\frac{iD_{0}}{\omega}\approx0$ which has little effect on the plasmon frequency.

From the microscopic Ohm's law we have $\boldsymbol{j}=-i\hat{\boldsymbol{\sigma}}\boldsymbol{q}\phi$, then the unit vector of the current density is

\begin{equation}
\boldsymbol{j}_{e}=\frac{1}{\mathcal{N}}\left(\begin{array}{c}
\frac{iD_{0}q_{x}\omega-q_{y}\sigma_{H}\omega^{2}-D_{0}q_{z}\omega_{c}}{\omega^{2}-\omega_{c}^{2}}\\
q_{x}\sigma_{H}+iD_{0}q_{y}/\omega\\
\frac{iD_{0}q_{z}\omega+D_{0}q_{x}\omega_{c}+iq_{y}\sigma_{H}\omega\omega_{c}}{\omega^{2}-\omega_{c}^{2}}
\end{array}\right),
\end{equation}
where $\mathcal{N}$ is the normalization coefficient. The functional form of the current density is universal and independent on the details of the plasmon dispersion. The Berry curvature of the plasmon is defined by \Citep{D. Jin 2016,J. C. Song 2018}

\begin{equation}
\boldsymbol{\Omega}(\boldsymbol{q})=-i\boldsymbol{\nabla_{q}}\times<\boldsymbol{j}_{e}|\boldsymbol{\nabla_{q}}|\boldsymbol{j}_{e}>.
\end{equation}
The Chern number of a plane in the reciprocal space reads as a integral that $C_{\alpha\beta}(q_{\gamma})=\frac{1}{2\pi}\int d\boldsymbol{S}_{\alpha\beta}\cdot\boldsymbol{\Omega}(q_{\alpha},q_{\beta},q_{\gamma})$. When $\omega_{c}=0$ or $\sigma_{H}=0$ or $D_{0}=0$, we have $\boldsymbol{\Omega}(\boldsymbol{q})=0$. It is a trivial phase. But when $D_{0}\sigma_{H}\omega_{c}\neq0$ and $D_{0}\rightarrow0$, $\boldsymbol{\Omega}(\boldsymbol{q})$ is diverging near $\boldsymbol{q}=0$. As a result, there emerges nonzero Chern numbers $C_{zx}=-1(C_{xy}=1)$ for a fixed $q_{y}=q_{0}(q_{z}=q_{0})$ where $q_{0}$ is an arbitrary nonzero real number. When there is a section of the body, corresponding topologically protected surface states will emerge. Just as shown in Fig.1(b), one could clearly see that the nontrivial geometric phase of the bulk plasmon is in consistent with the one-way propagation of the surface plasmons. On the other way, $C_{yz}$ is always zero and accordingly there are no unidirectional plasmons on the $z\ (y)$ direction of the top (front) surface, seen in Fig.S2 of \citep{Supplemental}.

$Semiclassical\ picture\ for\ Weyl\ orbits.$---In the presence of a uniform magnetic field $\boldsymbol{B}=(0,B,0)$, the equations of motion for an electron wavepacket in the Fermi arcs are \citep{D. Xiao 2010}

\begin{equation}
\dot{\boldsymbol{r}}=\frac{\partial E_{\boldsymbol{k}}}{\hbar\partial\boldsymbol{k}}-\dot{\boldsymbol{k}}\times\boldsymbol{\Omega}^{arc}(\boldsymbol{k}),
\end{equation}

\begin{equation}
\hbar\dot{\boldsymbol{k}}=-e\dot{\boldsymbol{r}}\times\boldsymbol{B},
\end{equation}
where $\boldsymbol{\Omega}^{arc}(\boldsymbol{k})=(\Omega_{yz},\Omega_{zx},\Omega_{xy})$ is the pseudovector of the Fermi-arc Berry curvature. Different from the traditional 2D electron gas, the electrons in Fermi arcs not only float on the surface, but also penetrate deep into the interior, forming a special 3D charge distribution. By replacing $k_{y}$ with $-i\partial_{y}$, we calculate the Fermi-arc Berry curvature $\boldsymbol{\Omega}^{arc}(\boldsymbol{k})$ and find that $\Omega_{zx}=0$, $\Omega_{xy}(\lambda k_{z},\lambda^{\prime}k_{x})=\Omega_{xy}(k_{z},k_{x})$, $\Omega_{yz}(\lambda k_{z},\lambda^{\prime}k_{x})=\lambda\lambda^{\prime}\Omega_{yz}(k_{z},k_{x})$ where $\lambda,\lambda^{\prime}=\pm1$ (see Sec. S5 of \citep{Supplemental}). Combining Eq.(11-12), one can see that the Fermi-arc electrons earn an anomalous velocity component perpendicular to the surface:

\begin{equation}
v_{\perp}=\frac{eB}{\hbar}(v_{z}\Omega_{xy}+v_{x}\Omega_{yz}).
\end{equation}

In Fig.2(b), we plot the variation curves of $\Omega_{yz}$ and $\Omega_{xy}$ along the top-surface Fermi arc. It indicates that near the Weyl nodes, the Berry curvature earns a sharp peak. So the electrons near Weyl nodes suffer the biggest impact and gain largest tunneling velocity, which is one order of magnitude bigger than the intrinsic speed. Suppose the magnetic field intensity $B$ is 1T, the relaxation time of electrons is 1$\mathrm{ps}$, the tunneling distance can reach 1 $\mathrm{\mu m}$, which could even be enhanced linearly by $B$. Such long mean free path can allow electrons to tunnel between the opposite surfaces without scattering \cite{X. C. Xie 2017,C. Zhang 2019}.

\begin{figure}
\begin{centering}
\ \ \ \ \ \ \ \ \includegraphics[scale=0.6]{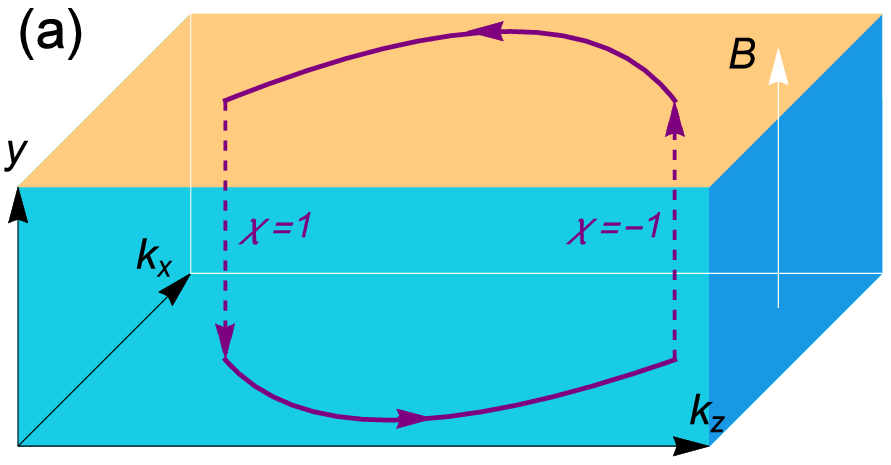}
\par\end{centering}
\begin{centering}
\ \
\par\end{centering}
\centering{}\includegraphics[scale=0.6]{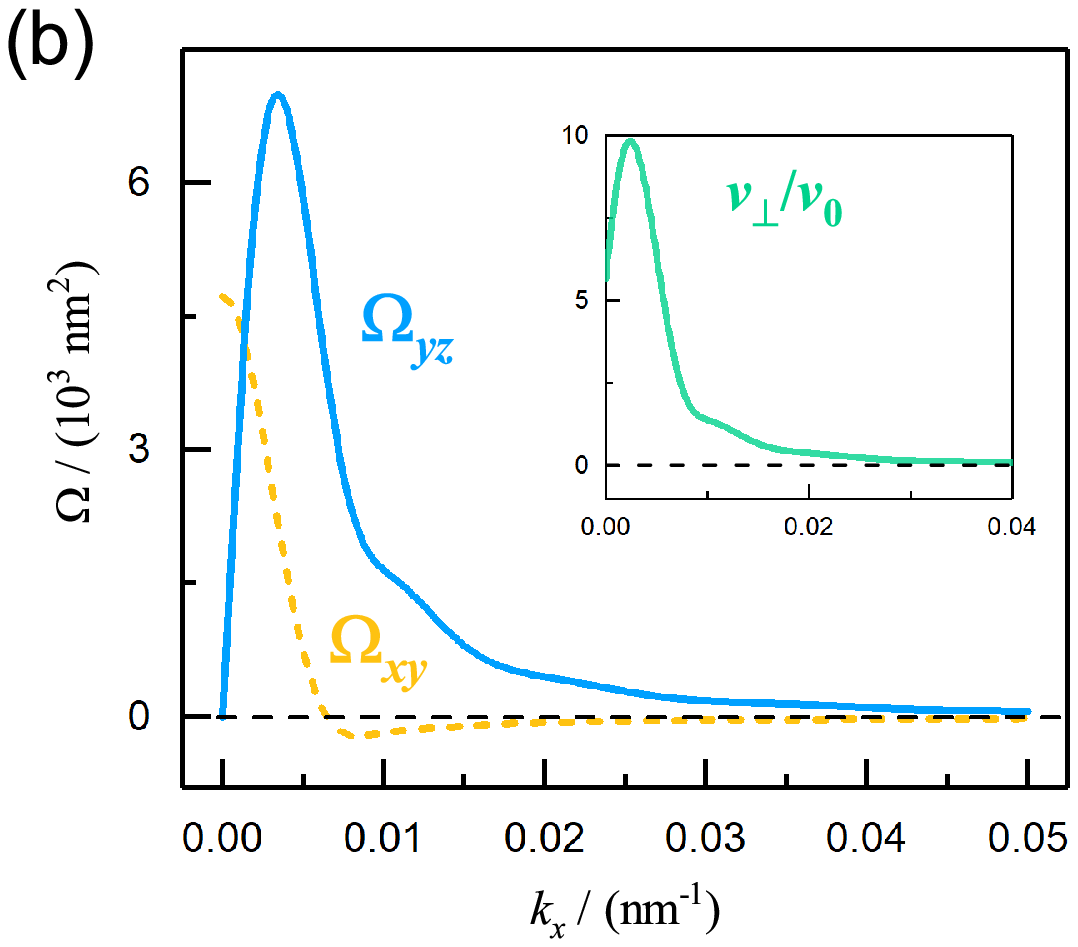}\caption{(a) A schematic diagram of the Weyl orbits in a slab of Weyl semimetal with thickness $L$. (b) Variation curves of the Berry curvature $\Omega_{xy}$ and $\Omega_{yz}$ along the top-surface Fermi arc for $k_{z},k_{x}>0$. The right inset shows the corresponding anomalous velocity perpendicular to the surface. $v_{0}=\sqrt{v_{z}^{2}+v_{x}^{2}}$. $A=0.5$eV$\cdot$nm, $M=5$eV$\cdot\mathrm{nm^{2}}$, $D_{1}=2$eV$\cdot\mathrm{nm^{2}}$, $D_{2}=3$eV$\cdot\mathrm{nm^{2}}$, $b=0.3\mathrm{nm^{-1}}$, $L=100\mathrm{nm}$, $B=1$T.}
\end{figure}

As $v_{z}(\lambda k_{z},\lambda^{\prime}k_{x})=\lambda v_{z}(k_{z},k_{x})$ and $v_{x}(\lambda k_{z},\lambda^{\prime}k_{x})=\lambda^{\prime}v_{x}(k_{z},k_{x})$, the anomalous velocity obeys $v_{\perp}(\lambda k_{z},\lambda^{\prime}k_{x})=\lambda v_{\perp}(k_{z},k_{x})$. Thus, the electron earns opposite motions near different Weyl nodes, resulting a complete cyclotron motion through the Weyl semimetal, just like Fig.2(a) shows.

Below we will show that when a top-surface plasmon reaches the boundary perpendicular to the $x$ axis, it would be reflected and tunnel to the bottom surface. As a quantum view, plasmons are the collective excitations of electron density oscillations. Before the reflection, the creation operator of the electron density reads $\boldsymbol{\rho}_{q_{z},q_{x}}^{+}=\underset{k_{z},k_{x}}{\sum}\boldsymbol{C}_{k_{z}+q_{z},k_{x}+q_{x}}^{+}\boldsymbol{C}_{k_{z},k_{x}}$ where $k_{x},q_{x}>0$. During the reflection, an electron with momentum $\boldsymbol{k}=(k_{z},k_{x})$ in the top Fermi arc will tunnel to the bottom through the Weyl node $\chi=1$ ending with a symmetric momentum $(k_{z},-k_{x})$. As a result, we get a creation operator of the electron density: $\boldsymbol{\rho}_{q_{z},-q_{x}}^{+}=\underset{k_{z},-k_{x}}{\sum}\boldsymbol{C}_{k_{z}+q_{z},-(k_{x}+q_{x})}^{+}\boldsymbol{C}_{k_{z},-k_{x}}$ which describes a fluctuation of the electrons at the opposite surface. So because of the ``wormhole'' tunneling through Weyl nodes and the surface-momentum lock, the top Fermi-arc plasmon will transform into a bottom one after the reflection.

$Topological\ Fermi\ arc\ plasmons.$---As a result of the formation of Weyl orbits, the Fermi-arc plasmons over opposite surfaces can make up an unique 3D topological plasmon. We first focus our attention on the top surface where $y=0$ and $q_{x}>0$. From the microscopic Ohm's law we have $\boldsymbol{j}^{s}=-i\boldsymbol{\sigma}^{s}\boldsymbol{q}\phi|_{0^{-}}$, then the unit vector of the surface current density is

\begin{equation}
\boldsymbol{j}_{e}^{s}=\frac{1}{\mathcal{N}_{s}}\left(\begin{array}{c}
\gamma q_{z}-i\eta q_{x}\\
q_{x}+i\gamma\eta q_{z}
\end{array}\right),
\end{equation}
where $\mathcal{N}_{s}=\sqrt{(1+\eta^{2})(q_{x}^{2}+\gamma^{2}q_{z}^{2})}$ is the normalization coefficient and $\gamma=\frac{\sigma_{zz}^{s}}{\sigma_{xx}^{s}}=\frac{D_{z}}{D_{x}}$. The functional form of Eq.(14) is universal and not dependent on the details of the plasmon dispersion. We define the pseudospin of the current density as $\boldsymbol{s}=<\boldsymbol{j}_{e}^{s}|\boldsymbol{\sigma}^{s}|\boldsymbol{j}_{e}^{s}>$:

\begin{equation}
s_{y}=-sin\theta,s_{z}=cos\theta cos\phi,s_{x}=cos\theta sin\phi,
\end{equation}
where $tan\theta/2=\eta$, $tan\phi/2=\gamma q_{z}/q_{x}$. Because $\eta\in[0,1]$, $\gamma q_{z}/q_{x}\in(-\infty,+\infty)$, one can see that $\theta$ changes from zero to $\pi/2$ and $\phi$ from $-\pi$ to $\pi$. When $\omega_{c}=0$, $s_{y}=0$, Eq.(15) describes a circle in the $s_{z}-s_{x}$ plane and there is no curvature in it. But when $\omega_{c}>0$, $s_{y}<0$, the pseudospin $\boldsymbol{s}$ will distribute on a semisphere exhibiting a curved pseudospin texture. As a result, there emerges a nonzero geometric phase in the Fermi-arc plasmons.

The electrodynamic equations can be transformed into an equivalent Hamiltonian eigenvalue problem $\omega\boldsymbol{j}^{s}=H\boldsymbol{j}^{s}$ where (see Sec. S2.7 of \citep{Supplemental}):

\begin{equation}
H=\frac{\boldsymbol{\sigma}^{s}}{i[(\epsilon+1)\varepsilon_{0}q-\sigma_{H}q_{x}/\omega]}\left(\begin{array}{cc}
q_{z}^{2} & q_{z}q_{x}\\
q_{x}q_{z} & q_{x}^{2}
\end{array}\right).
\end{equation}
One can find that the eigenvector of the Hamiltonian is exactly Eq.(14). When $\sigma_{H}=0$, Eq.(16) describes the traditional 2DEG magnetoplasmon problem \Citep{D. Jin 2016,J. C. Song 2018}. When $q_{x}>0$ for the top surface, straightforward derivation implies the Chern number $C_{top}$ is $\frac{1}{2}$ (see Sec. S2.6 of \citep{Supplemental}). Likewise, for the bottom we have $C_{bott}=\frac{1}{2}$. Therefore there is a nontrivial 3D topological plasmon with Chern number $C=1$ over the opposite Fermi arcs.

\begin{figure}
\noindent \begin{centering}
\includegraphics[scale=0.7]{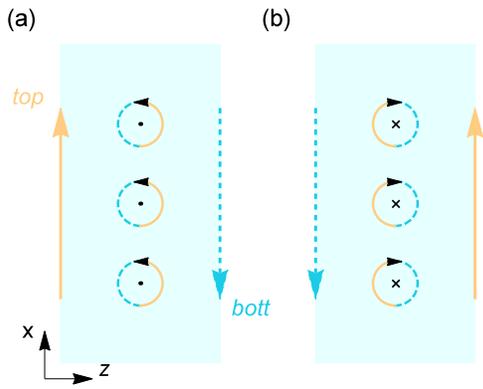}
\par\end{centering}
\caption{Diagrammatic top view of the direction and position of the unidirectional edge plasmons with different magnetic fields. The circles with arrows denote the cyclotron motion of the Fermi-arc electrons.}
\end{figure}

$Unidirectional\ edge\ plasmons.$---Now we come to consider the edge states of the topological Fermi-arc plasmons. Near the left edge as plotted in Fig.3(a), we assume that the electric potential of the edge plasmons takes a form as $\phi(z,x)=\phi_{0}e^{-k_{z}z+iq_{x}x}$ where $k_{z}>0$, $z>0$. Compared with the surface plasmons where $\phi\propto e^{iq_{z}z+iq_{x}x}$, the edge plasmon can be described with Eq.(14) by replacing $q_{z}$ with $ik_{z}$. From the boundary condition $j_{z}^{edge}|_{0^{+}}=0$, we obtain $k_{z}=\frac{\eta q_{x}}{\gamma}$. Because $\eta,\gamma>0$, one can find that $q_{x}>0$ and so the edge plasmon is unidirectional propagating along the positive $x$-axis. On the contrary, for the right edge the edge plasmon goes along the negative direction.

When the external magnetic field is along the negative direction of the $y$-axis, as plotted in Fig.3(b), the cyclotron motion of the Fermi-arc electrons will turn around and $\eta$ becomes negative. Accordingly, the top edge plasmon would propagate along the right boundary and the bottom along the left. Therefore, the direction and position of the unidirectional edge plasmons can be adjusted by the external magnetic field.

In the low frequency range $\omega\ll\alpha\tilde{\sigma}_{H}$, combining Eq.(14) and Eq.(16) one can get the frequency of the edge plasmons:

\begin{equation}
\omega_{edge}=(v-\frac{D_{x}}{\tilde{\sigma}_{H}})q_{x},
\end{equation}
which implies a gapless linear edge mode independent of the external magnetic field.

In summary, we have pointed out that there are two kinds of 3D topological plasmons in presence of a magnetic field: the bulk plasmons and the Fermi-arc plasmons over opposite surfaces. According to the bulk-boundary correspondence, there are unidirectional surface/edge plasmons whose direction and dispersion can be controlled by the external field. These chiral surface plasmons possess momentum-location lock and can tunnel to the opposite surface. The anomalous Hall conductivity can greatly change the magnetoplasmon dispersion and gives rise to linear, parabolic or even hyperbolic bands. Strong confinement of the EM field associated with Fermi-arc plasmons has been found. In addition, a semiclassical picture of electron motion is proposed to show the formation of Weyl orbits and the influence of ``wormhole'' tunneling on plasmon transport. Our work thus provides instructive insights into the electron dynamics and collective excitations of Weyl fermions and suggests Weyl semimetal a good seed for 3D topological plasmonics.

This work was supported by National Key Research and Development Program of China (Grant No. 2017YFA0303400), NSFC-RGC (Grant No. 11861161002) and National Natural Science Foundation of China (Grant No. 11774036).

\end{document}